\documentclass[journal=chemmater,manuscript=article,layout=twocolumn]{achemso}

\usepackage[version=3]{mhchem} 
\usepackage{hyperref}
\usepackage{color}

\author{Prabuddha Kant Mishra}
\affiliation{Department of Chemistry, Indian Institute of Technology Delhi, New Delhi 110016, India}
\author{Shivani Kumawat}
\affiliation{Department of Physics, Indian Institute of Technology Delhi, New Delhi 110016, India}
\author{Soumyakanta Panda}
\affiliation{School of Basic Sciences, Indian Institute of Technology Bhubaneshwar, Odisha 752050, India}
\author{Niharika Mohapatra}
\affiliation{School of Basic Sciences, Indian Institute of Technology Bhubaneshwar, Odisha 752050, India}
\author{B K Mani}
\affiliation{Department of Physics, Indian Institute of Technology Delhi, New Delhi 110016, India}
\author{Ashok Kumar Ganguli}
\email{ashok@chemistry.iitd.ac.in}
\affiliation{Department of Chemistry, Indian Institute of Technology Delhi, New Delhi 110016, India}
\alsoaffiliation{Department of Chemical Sciences, Indian Institute of Science Education and Research, Berhampur, Odisha-760003, India}

\title[An \textsf{achemso} demo]{Antiferromagnetic weak topological state in Bismuth square-net based nonsymmorphic lattice}

\abbreviations{RWR, DM}

\begin{document}


\begin{tocentry}

\includegraphics[width= 1.0\columnwidth,angle=0,clip=true]{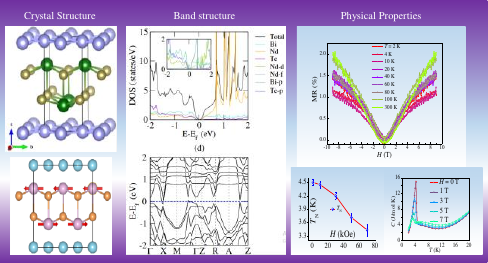}

\end{tocentry}

\begin{abstract}

        The ZrSiS-class of layered materials offer interesting topological and magnetic 
	characteristics suitable for spintronics applications. In this work, 
	we have synthesized a polycrystalline 
	NdBiTe using solid-state reaction technique and have examined the magnetic 
	properties in 2 - 300 K temperature range using temperature and field-dependent 
	magnetization measurements. Our magnetic and specific heat data
	demonstrates a long-range antiferromagnetic ordering in the material below 4.5 K. 
	Furthermore, our isothermal magnetization data show a signature of
	spin-reorientation below Neel temperature. 
	The observed nonlinearity in inverse susceptibility vs temperature data, and a 
	hump in specific heat in 5-20 K range, indicate the existence of 
	crystal field splitting in the material.  Our transport properties 
	measurements show the metallic behavior with positive 
	magnetoresistance in the temperature range of 2 - 300 K. The observed 
	rise in resistivity as function of temperature below Neel temperature infers 
	the strongly correlated fermions, which is consistent with the 
	observed large Sommerfeld coefficient. Consistent with experimental results, 
	our first-principles calculations predict an antiferromagnetic semimetallic 
	nature of NdBiTe. Further, our spin-orbit coupled simulations of electronic 
	structure show a signature of weak topological nature of the material. 
	
\end{abstract}



\maketitle

\section{Introduction}

Systematic understanding of the interplay among topology, magnetism, and electrical 
properties of materials 
has been of prime interest for the last few decades. Materials with correlated 
electronic interactions play a pivotal role in comprehending the mechanistic 
links and inter-dependency between various degrees of freedom like spin and charge. 
Rare-earth rich systems are prone to have magnetic interactions due to the presence 
of $f$-electrons and the properties of such materials are substantially influenced 
by interactions among conducting electrons and localized moments, termed as RKKY 
(Ruderman-Kittel-Kasuya-Yosida) interaction. A family of ZrSiS-type structures 
has been explored for topological and magnetic 
properties \cite{Sankar2020, Pandey2020, Chen2017, Regmi2022, Gao2022, Yang2020}. 
Layered materials often exhibit crystal-field splittings due to the reduced 
symmetry arising from stacking along a particular axis.  The interlayer and 
the inter-ligand distances have been found to be crucial to affect the extent 
of crystal-field energy splitting. In addition to it, both theoretical and 
experimental investigations have been performed to reveal the presence of 
topological properties intricately connected with non-symmorphic layered 
structures in these materials.  The degeneracy of nodal-line in nonsymmorphic 
materials is protected by crystal symmetry which makes it different from 
Dirac semimetals and Weyl semimetals \cite{Schoop2018}.
Such intertwined aspects of structure and topology of the electronic band 
have been explored for nodal-line semimetal ZrSiX (X = S, Se, and Te) by 
ARPES (angle-resolved photoemission spectroscopy) and magnetotransport 
measurements \cite{Neupane2016, Topp2016, Schoop2016, Fu2019, Ali2016, Singha2017, Song2022}.

The materials consisting of the square-net like structures have been 
investigated to have symmetry-protected four-fold Dirac states in Brillouin 
zone \cite{Klemenz2020, Klemenz2019}. The theoretical predictions identified 
LaSbTe, a nonmagnetic material, as a potential weak topological insulator due 
to its significant spin-orbit coupling \cite{Qiunan2015}. 
Subsequent experimental validation confirmed the presence of a Dirac-like 
dispersion within its electronic band structure.\cite{Singha2017a}. 
In the case of magnetic isostructural analogues, magnetism is an additional 
tuning parameter to be explored. 
Non-trivial topological phase in long-range antiferromagnetically ordered 
materials has been observed through both experiments as well as first-principles 
simulations. Experimentally it was observed in ARPES measurement in GdSbTe \cite{Hosen2018} 
and CeSbTe \cite{Schoop2018}, whereas theoretically it was reported 
in HoSbTe \cite{Yang2020}. Diverse magnetic features 
like devil's staircase and spin-flop transitions have been observed 
in various members of LnSbTe (Ln = rare earth). Moreover, the coexistence of 
competitive short-range magnetism and long-range ordering has been explored 
in HoSbTe, TbSbTe \cite{Plokhikh2022} and CeSbTe \cite{Chen2017} materials. 
Interestingly, the antiferromagnetically ordered magnetic moments exhibit 
a change in their orientations relative to the square lattice arrangements of 
the rare-earth elements in ZrSiS materials. This suggests a change in the interatomic 
exchange interaction, allowing a tunable single-ion anisotropy. 
Though ZrSiS-class of LnSbTe are known and have been studied for 
physical properties, Bi-square net based ZrSiS-type of materials are still unexplored. 
Such materials are however important as they exhibit exotic fundamental 
as well as functional properties due to their larger atomic 
radii and stronger spin-orbit coupling compared to their Sb-analogues.

In this study, we have successfully synthesized the polycrystalline NdBiTe, 
a new member of the ZrSiS family, by solid-state reaction technique.
NdBiTe is a bismuth-based layered magnetic material, which requires 
a focused attention to comprehensively investigate the interplay among 
magnetism, nontrivial topological characteristics, and transport properties.
Detailed magnetic measurements have been carried out to explore 
the nature of various interactions. 
With the help of 
temperature-dependent and isothermal magnetic measurements, our 
study found a robust long-rang antiferro-type magnetic ordering in NdBiTe. 
Furthermore, the field-dependent specific heat studies infer details of 
thermal parameters and establish it as a strongly correlated electronic 
system. Due to the effect of strongly correlated electrons, resistivity rises 
below Neel temperature. 
The observed nontrivial field dependence of MR reveals its genesis 
from the nontrivial topology of the electronic band structure.
In addition, to support our experimental results, we performed 
first-principles density functional theory calculations of 
the relevant properties in NdBiTe. Our computed results are 
consistent with our measured values.

The remainder of the paper is organized in two sections. In Section II, we
provide brief details on experimental and computational methods used in the work. 
In Section III, we discuss and analyze the results
from our experiment and calculations. For an easy discussion, we have
separated this section in to five subsections.

\begin{figure*}
\begin{center}
\includegraphics[width= 1.5\columnwidth,angle=0,clip=true]{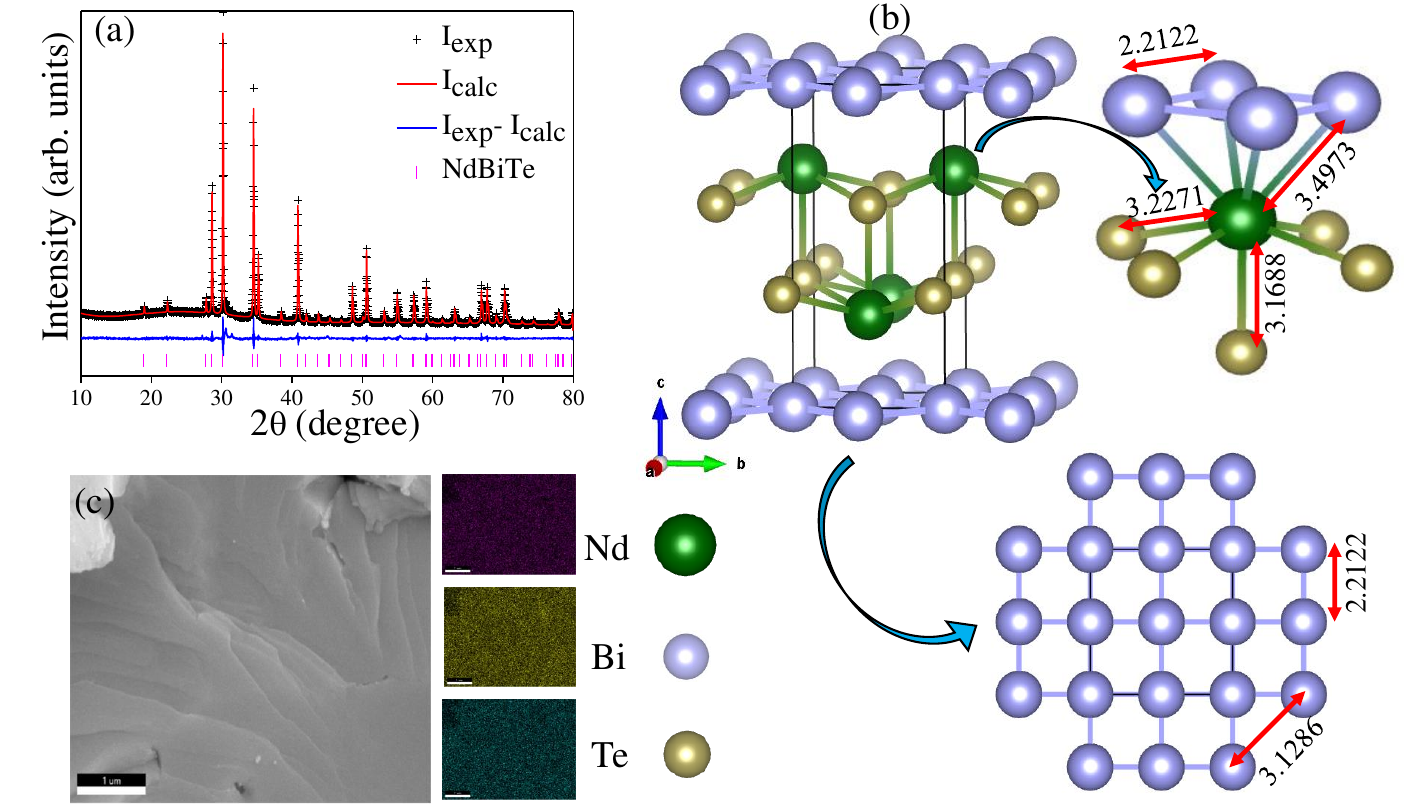}
\caption{(Color online) (a) Rietveld refinement of room temperature 
	powder X-ray diffraction data of polycrystalline NdBiTe. Vertical 
	bars show the allowed Bragg's peak for NdBiTe. The blue line represents 
	the difference in the experimental and simulated patterns. 
	(b) Shows the Crystal structure of NdBiTe, with constituents of 
	polyhedra around Nd ion and the square-net layer of Bismuth, 
	mentioned with various interatomic distances in \r{A} unit. 
	(c) The SEM image shows the layered nature of the specimen along with
	elemental mapping to show the homogeneous distribution of elements.}
\label{fig1}
\end{center}
\end{figure*}

\section{Experimental and Computational Details}

Polycrystalline NdBiTe has been synthesized by solid-state reaction method 
using a sealed tube technique. The elemental form of Nd, Bi, and Te have 
been ground together in the stoichiometric ratio (Nd:Bi:Te::1:1:1) in an 
argon-filled glove box and sealed in an evacuated quartz tube for heat 
treatment in a programmed high-temperature furnace. A heating condition 
of 973 K for 48 hours was used for the reaction. The product of the first 
heating was ground well, pelletized, and sintered at 973 K for 48 hours 
for better phase homogeneity. Resulting phases have been characterized 
using X-ray diffraction technique using Bruker D8 Advance diffractometer 
with Cu-K$\alpha$ radiation performed on powder, shown in Fig. \ref{fig1}. 
The structural refinement on powder X-ray diffraction data was carried 
out using the Rietveld method with the TOPAS software package.
By using the EDX (energy dispersive X-ray analysis) technique, we confirmed the
homogeneity and composition of the sample. The temperature and
field-dependent magnetic measurements were performed in a superconducting
quantum interference device (MPMS-3). The specific heat and electrical transport
were measured as a function of temperature and applied field by 
using physical property measurement system (PPMS, Quantum Design).

\begin{table*}
\scriptsize\addtolength{\tabcolsep}{-1pt}
\caption{\label{table1} Structural and magnetic parameters of 
	NdBiTe in their respective units.}
\centering
\begin{tabular}{l c c c c r}
\hline
\hline
Space group: & $P4/nmm$ \\
Space group number: & 129 \\
{\it a} (\r{A}): & 4.4245(1) \\
{\it c} (\r{A}): & 9.3775(2) \\
\hline
Atom & Site & x    & y    & z         &  B$_{\rm iso}$ \\
  Nd & 2c   & 0 & 1/2 & 0.2889(3) & 1.13(6) \\
  Bi & 2a   & 0 & 0 & 0         & 1.33(5) \\
  Te & 2c   & 0 & 1/2 & 0.6268(3) & 2.19(7) \\
\hline
	$T_{\rm N}$ &&& \multicolumn{3}{r}{4.5(1) K}  \\
	$\Theta_{\rm P}$ &&& \multicolumn{3}{r}{-18.78(5) K}    \\
	$C$ &&& \multicolumn{3}{r}{1.652(2) emu Oe$^{-1}$mol$^{-1}$K$^{-1}$}  \\
	$\chi(0)$ &&& \multicolumn{3}{r}{0.00133 emu Oe$^{-1}$mol$^{-1}$}    \\
	$\mu_{\rm eff}^{\rm theo}$ &&& \multicolumn{3}{r}{3.63 $\mu_B$/Nd$^{+3}$}    \\
	$\mu_{\rm eff}^{\rm calc}$ &&& \multicolumn{3}{r}{3.68(5)$\mu_B$/Nd$^{+3}$}    \\
	$M_{\rm S}$(2K) &&& \multicolumn{3}{r}{1.16(2) $\mu_B$ mol$^{-1}$}    \\
	$H_{\rm C}$(2K) &&& \multicolumn{3}{r}{45(5) Oe}   \\
 \hline
 \hline
\end{tabular}
\end{table*}
\begin{figure*}
\begin{center}
\includegraphics[width=1.55 \columnwidth,angle=0,clip=true]{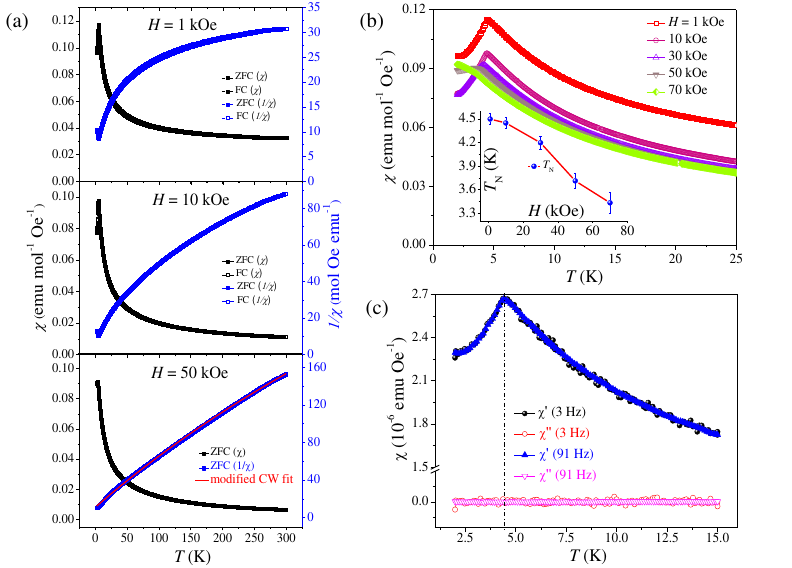}
	\caption{(Color online) (a) Magnetic susceptibility ($\chi(T)$) and corresponding 
	inverse susceptibility ($\chi^{-1}(T)$) in 
	zero-field-cooling (ZFC) and field-cooling (FC) at different fields (1, 10 
	and 50 kOe). $\chi^{-1}(T)$ at the 
	applied field of 50 kOe, fitted with modified C-W model, 
	shown by red solid line. (b) Variation of ZFC $\chi(T)$ at various
        applied magnetic fields in low-temperature regime (2-25 K),
	the inset shows the suppression of $T_N$ as a function of the applied field.
	(c) The $\chi'$ and $\chi''$ components of ac-susceptibility data in 
	the temperature range of 2-15 K at 3 and 91 Hz frequencies.}
\label{fig2}
\end{center}
\end{figure*}

To support our experimental results, we performed density functional 
theory (DFT) based first-principles calculations using the projected augmented 
wave (PAW) method \cite{Blochl1994} as implemented in the 
{\it Vienna Ab initio Simulation Package} (VASP) 
code \cite{Kresse1999, Kresse1994, Kresse1996}. 
The exchange-correlations among electrons were incorporated using the 
generalized gradient approximation (GGA) based Perdew-Burke-Ernzerhof 
(PBE) pseudopotential \cite{Perdew1996}. The plane-wave 
basis with an energy cutoff of 500 eV were used in all the calculations. 
The crystal structures were optimized using full relaxation calculations
with the help of a conjugate gradient algorithm with a $\Gamma$-centered 
k grid of 10$\times$10$\times$4. The energy and force convergence 
criteria of $10^{-6}$ and $10^{-5}$ eV, respectively, are used in 
all the self-consistent-field calculations. 
To incorporate the effect of strongly correlated $4f$-electrons of Nd, 
on-site Coulomb interaction was included using the rotationally invariant 
DFT+U approach of Dudarev et al.\cite{Dudarev1998, Dudarev1997}. 
Our calculated U value, 6.58 eV, is consistent with the previous 
calculations for similar systems \cite{Gebauer2021}. 
Further, to incorporate the relativistic effects,
arising from the heavier atoms Bi and Nd, we included spin-orbit 
coupling (SOC) in all the calculations.
A supercell of $2\times2\times2$ (48 atoms) was used to calculate 
the phonon dispersion and other relevant properties. The phonon 
dispersions were calculated with $\Gamma$-centered $k$-mesh of 
$3\times3\times2$ using linear response density functional 
perturbation theory (DFPT) as implemented in 
PHONOPY \cite{Dove1993, Togo2023}.

\section{RESULTS AND DISCUSSION}

Fig. \ref{fig1}(a) shows the Rietveld refinement of room temperature 
powder X-ray diffraction pattern of NdBiTe. It crystallizes in the ZrSiS-type 
centrosymmetric tetragonal structure with space group $P4/nmm$ and the lattice 
parameters are $a$ = 4.4245(1) \r{A} and $c$ = 9.3775(2) \r{A}. The details of 
refined structural parameters are given in Table \ref{table1}. 
Fig. \ref{fig1}(b) shows the crystal structure of NdBiTe,  where 
Nd and Te ions occupy the 2a sites while 2c sites are filled with Bi, 
and each Nd-ion is coordinated by {\it five} Te ions and {\it four} Bi ions. 
The polyhedral structure around Nd can be seen as a square antiprismatic geometry. 
This geometry will lead to anisotropic magnetism and crystal-field energy 
splitting for otherwise degenerate Nd-f electrons.
Like other ZrSiS materials, NdBiTe can be 
visualized as the stacking of two distinct layered components [NdTe] and Bi. 
The interatomic interaction will be helpful in deciding the effective bonding 
among these for compound formation. 
In earlier reports, the c/a ratio is used for comprehension of the nature of bonding in layered materials \cite{Klemenz2020}. For NdBiTe, the c/a is 2.12, which is close to that of GdBiTe (c/a = 2.11). Thus NdBiTe belong to same structural class of ZrSiS type (c/a =2.27) layered materials having ionic interaction among adjacent layers \cite{Gebauer2021}.
The delocalized bonding in the square-net of Bi is key to transport properties and stabilizes the sqaure-net geometry \cite{Klemenz2019, Klemenz2020, Schoop2018}. The intraplanar Bi-Bi bond length is found to be shorter in NdBiTe (3.129 \r{A}) as comparable to its structural analogue GdBiTe (3.0905\r{A}). Here, it is important to mention that the observed Bi-Bi bond length is comparable to the puckered honeycomb units in the elemental form of bismuth (3.071\r{A}) \cite{Cucka1962}. In [NdTe] layer, two distinct Nd-Te bond lengths are observed as shown in the Fig. \ref{fig1}(b). The four in-plane ($\parallel$ ab plane) Nd-Te bonds are longer (3.227\r{A}) than one out-of-plane Nd-Te bond (3.169\r{A}) i.e. $\parallel$ to c-axis.
\begin{figure*}
\begin{center}
\includegraphics[width=2 \columnwidth,angle=0,clip=true]{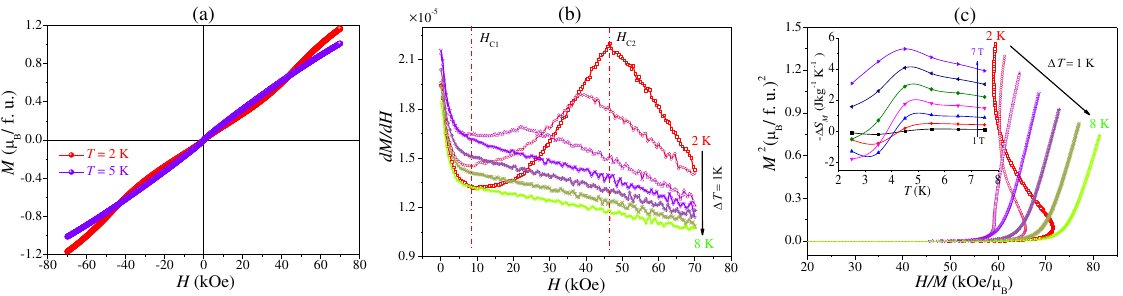}
\caption{(Color online) (a) Isothermal magnetization ($MH$) data as a function
	of applied magnetic field at 2 and 5 K. 
	(b) $dM/dH$ as a function of applied magnetic field at various temperatures 
	across the magnetic transition from 2 K to 8 K, with $\Delta T = 1$ K. 
	The dotted red lines show the critical fields ($H_{\rm C1}$ and $H_{\rm C2}$) 
	for 2 K curve, as a case of representation. 
	(c) The corresponding Arrott plots and inset demonstrates the change in 
	entropy of magnetization as a function of temperature for various fields 
	ranging from 10 kOe to 70 kOe.}
\label{fig3}
\end{center}
\end{figure*}

\subsection{Temperature Dependent Magnetic Susceptibility}

Fig. \ref{fig2}(a) shows the temperature-dependent magnetization 
measurements where we have shown our data on $\chi$ (T) at different applied 
fields along with inverse susceptibility in the temperature range of 2-300 K, 
measured in both zero-field-cooled (ZFC) and field-cooled (FC) protocols. 
As discernible from the panel (a), the $\chi(T)$ data with ZFC and FC protocols 
overlap with each other. 
This suggests the absence of irreversibility through competing magnetic 
interactions or glassy-spin nature in the material \cite{Zhang2017}.
The magnetization increases with a decrease in the temperature up to the 
Neel temperature ($T_N$ = 5 K) and below $T_N$ it decreases due to the 
antiferromagnetic (AFM) alignment of the magnetic spins. 
The inverse susceptibility versus temperature data show a significant 
deviation from the linear behavior and does not follow the Curie-Weiss (C-W) law 
for a larger range of low temperatures. 
The reason for this nonlinear behavior of $\chi^{-1}\;vs\;T$ could be attributed 
to the crystal field splitting of the energy levels, which gets suppressed 
gradually as the applied field increases from 1 kOe to 50 kOe. A similar trend 
of deviation from C-W fit below 150 K is reported in a single crystal of CeSbTe \cite{Chen2017}. 
From the modified C-W fit, $C$ is estimated to be 1.652 (2) emu K Oe$^{-1}$ 
mol$^{-1}$ and $\theta_p$ (Weiss-temperature) is obtained as -18.78 K, 
which suggests an antiferromagnetic-type exchange interactions between the spins. 
The $\chi_0$ was estimated to be 0.00133(1) emu Oe$^{-1}$ mol$^{-1}$, which was used further 
to calculate the effective magnetic moments ($\mu_{\rm eff}$) of the material. 
The $\mu_{\rm eff}$ was extracted to be 3.635 $\mu_{\rm B}/{\rm Nd}^{+3}$ ions, 
which is consistent with theoretical value ($g\sqrt{J(J+1)}\mu_B$) 3.63 
$\mu_{\rm B}/{\rm Nd}^{+3}$.

Next, as the sharpness of transition at $T_{\rm N}$ decreases with increase in 
the applied field, we carried out the first derivative of $\chi(T)$ at different 
fields with respect to temperature to determine the Neel temperature. As can be 
observed from the panel (b) of the figure, the paramagnetic (PM) to AFM 
transition is robust for applied external magnetic fields and the $T_{\rm N}$ 
is found to decrease gradually with an increase in the field strengths, 
eventually reaching to $\approx$ 3.4 K for $H$ = 70 kOe 
(inset of Fig. \ref{fig2}(b)). It is worth mentioning that the 
AFM long-range orderings have been observed for the isostructural magnetic 
materials with rare-earth elements \cite{Sankar2020, Pandey2020}. 
Furthermore, to resolve the possibility of the existence of glassy spin 
interaction at low-temperature regime, {\it ac}-susceptibility measurements were 
performed at 3 and 91 Hz frequencies. As discernible from the Fig. \ref{fig2}(c), 
the {\it ac}-susceptibility data show 
a peak-shaped anomaly close to 4.48 K ($T_{\rm f}$) for the in-phase component 
of susceptibility ($\chi'$), with no signature of frequency dependent behavior.
In addition, the out-of-phase component ($\chi''$) shows no significant change 
as a function of temperature, which suggests the absence of spin-glass-like 
short range interactions. 

\begin{figure*}
\begin{center}
\includegraphics[width= 2\columnwidth,angle=0,clip=true]{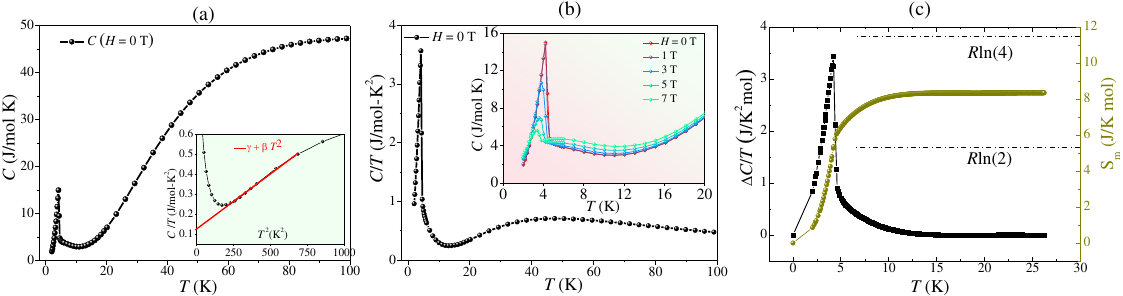}
\caption{(Color online) (a) The specific heat data as a function of temperature 
	in the absence of applied magnetic field. The inset shows $C/T$ $vs.$ $T^2$ 
	plot fit (solid red line) with equation $\gamma + \beta \rm{T}^2$. 
	(b) $C/T$ $vs.$ $T$ plot signifying the peak centered at $\sim$ 48 K, 
	corresponding to Schottky anomaly. The inset shows a temperature-dependent 
	specific heat at various applied magnetic fields.}
\label{fig4}
\end{center}
\end{figure*}
\subsection{Isothermal Magnetization}

Next, we performed isothermal $M(H)$ measurements to probe magnetization 
in the vicinity of $T_{\rm N}$, as shown in Fig. \ref{fig3}. 
As discernible from the panel (a), $M-H$ loops are consistent with the 
antiferromagnetic nature of NdBiTe below $T_{\rm N}$. The $M(H)$ data at 
2 K has a change in the slope with an increase in the magnetic field, suggesting 
a spin-reorientation in the material.  The change in the slope of $M_{\rm H}$ curves 
is more visible in the $dM/dH$ plots shown in panel (b)) at various temperatures. 
There are two points of inflection in the $dM/dH$ vs $H$ data, denoted as 
$H_{\rm c1}$ and $H_{\rm c2}$ to represent the corresponding critical 
fields for spin reorientation. As can be observed from the figure, 
the $H_{\rm c2}$ decreases gradually from 46.5 kOe to 21.5 kOe with an 
increase in the temperature from 2 K to 4 K, and eventually disappears above 
the $T_{\rm N}$. This suggests the weakening of the antiferromagnetic interaction 
among spins with increase in the temperature. 
Contrary to $H_{\rm c2}$, there is no significant change in $H_{\rm c1}$ 
with temperature. 
The change in the slope with magnetic fields suggests a transition from AFM 
state to a canted-AFM state, where the canting of spins leads to a deviation 
from the linearity in isotherms and adds a ferromagnetic-like response 
in the magnetization data. The presence of spin-reorientation-driven 
metamagnetic transitions in isothermal magnetization at low temperatures 
is also observed in other isostructural materials with AFM 
ordering \cite{Pandey2020, Schoop2018, lv2019, Lei2019, Gao2022}. 
The saturation magnetization observed from our experiment is 
1.18 $\mu_{\rm B}/{\rm Nd}^{+3}$ at 2 K with an applied field of 7 T.
This is comparable to the observed average value of magnetization for NdSbTe
by Pandey et al. ($1 \mu_{\rm B}/{\rm Nd}^{+3}$) \cite{Pandey2020}, 
and Sankar et al. ($1.25 \mu_{\rm B}/{\rm Nd}^{+3}$) \cite{Sankar2020}.
The relatively large value obtained by Sankar et al., has been attributed to anisotropic magnetization.
The observed value of saturation magnetization is smaller in comparison 
to the theoretical value 3.27 $\mu_{\rm B}/{\rm Nd}^{+3}$, indicating the presence
of appreciable single-ion anisotropy \cite{Sankar2020, Regmi2023}. 
Next, we used the virgin plots, recorded at various temperatures around the 
$T_{\rm N}$, to extract the Arrott plot. This is shown in the
panel (c) of the figure. The negative slope of curves in the Arrott plot 
suggests the first order of phase transition to AFM state, supported by 
Banerjee criterion \cite{Banerjee1964}.

Further, the change in the magnetization entropy, $\Delta S_{\rm M}$, 
associated with the magnetic transition was obtained from the 
pyromagentic coefficient $\frac{\partial M (T, H)}{\partial T}$ 
using the relation
\begin{equation}
  \Delta S (T, H_0)=  \int^{H_0}_0 \left[ \frac{\partial M (T, H)} 
	              {\partial T} \right]_{H} dH.
  \label{entropy2}
\end{equation}
The $\Delta S_{\rm M}$ as a function of temperature at various magnetic 
fields is shown in the inset of Fig. \ref{fig3} (c). As can be observed 
from the data, it resembles the $M(T)$ behavior from dc susceptibility 
Fig. \ref{fig1}(a). 
$\Delta S_{\rm M}$ acquires a positive value at low applied field (0.1 T), 
suggesting the presence of antiferromagnetic interaction at low fields. 
The sign of $\Delta S_{\rm M}$ gradually changes from positive to negative 
with increasing magnetic field strengths. The negative value indicates 
the spin-reorientation for higher applied fields and has a maximum value 
$\approx$ -5.15 J kg$^{-1}$K$^{-1}$ at 4.5 K for $\Delta H$ = 7 T.

\subsection{Specific Heat}

To gain further insights into the magnetic interactions embedded in 
antiferromagnetic ordering at low temperatures,  we performed specific heat 
measurements at various applied magnetic fields. Fig. \ref{fig4}(a) shows 
the specific heat data in the absence of applied magnetic field over a 
range of temperatures. As discernible from the figure, we observed an anomaly 
at 4.5 K, confirming a long-range magnetic ordering. This is consistent with the 
temperature-dependent susceptibility behavior shown in Fig. \ref{fig2}. 
The field-dependent specific heat data from our measurement are shown in 
the inset of Fig. \ref{fig4}(b). As can be observed from the figure, the 
intensity of the anomaly decreases, and also shifts to lower temperatures, with 
increase in the strengths of applied magnetic field. This is a signature 
of an antiferromagnetic interaction. 

\begin{figure*}
\begin{center}
\includegraphics[width= 2 \columnwidth,angle=0,clip=true]{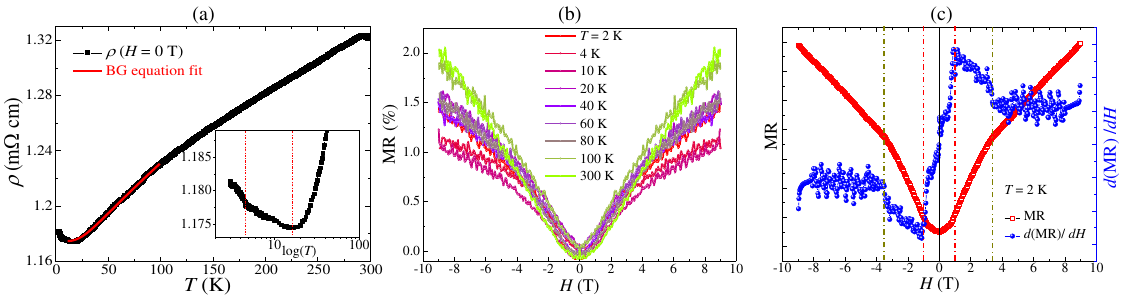}
\caption{(Color online) (a) Temperature-dependent resistivity $\rho(T)$ 
	in 2 - 300 K range. The solid red line shows a fit to BG equation 
	in the high-temperature regime. The inset shows $\rho(T)$ data in 
	logarithmic scale, where the dotted red lines represent an upturn 
	in $\rho(T)$ at $\approx$ 16 K and a slope change at $T_{\rm N}$. 
	(b) The MR data as a function of $H$ at different temperatures. 
	(c) The smoothed MR data as function of $H$ and its first derivative 
	with respect to the applied field. The dotted lines indicate 
	the critical fields.}
\label{fig5}
\end{center}
\end{figure*}

Since our measured specific heat data embeds the contributions from 
electronic, lattice, and magnetic interactions, to asses their individual 
contributions we fit the data with temperature-dependent specific 
heat equation, in the low-temperature regime, as
\begin{equation}\label{Debye}
	C = \gamma T + \beta T^3 + C_{\rm m}.
\end{equation}
Here, $\gamma$ and $\beta$ are the Sommerfeld and phonon coefficients, 
respectively, and $C_{\rm m}$ is the magnetic contribution to the specific 
heat.  The data well above $T_{\rm N}$, in the range of 18-26 K, can be 
fit without considering magnetic contribution \cite{Chen2017, Gao2022}. 
The $C/T$ $vs$ $T^2$ data shows 
a linear fit (shown with a solid red line in the inset of panel (a)), 
with the parameters $\gamma$ and $\beta$ estimated to be 126(4) mJ/K$^2$ 
mol and 0.55(1) mJ/K$^4$mol, respectively. The relatively larger value 
of Sommerfeld coefficient as compared to isostructural antiferromagnets 
(e.g., 115 mJ/K$^2$ mol for NdSbTe) \cite{Pandey2020} and much 
larger as compared to nonmagnetic materials (e.g., 2.19 and 0.51 mJ/K$^2$ 
mol for LaSbSe and LaSbTe, respectively \cite{Pandey2022}), suggests 
strong electronic correlations in NdBiTe due to the presence of 
Nd-$4f$-electrons. There is a possibility of correlated fermionic 
behavior, resulting in the mass enhancement of fermions. This is consistent with 
the findings reported in other isostructural magnetic systems \cite{Pandey2020}.

The value of phonon coefficient, $\beta$, was used further to estimate 
the Debye temperature using the relation $\theta_D= (12\pi^4NR/5\beta)^{1/3}$, 
where $N$ = 3 for NdBiTe and $R$ is the universal gas constant. 
The estimated value of $\theta_D$ = 147(1) K for NdBiTe is lower than 
that of NdSbTe ($\theta_{\rm D}$ = 236 K by Pandey et al\cite{Pandey2020} 
and 153 K by Sankar et al \cite{Sankar2020}), indicating weaker interatomic 
interactions present in NdBiTe as compared to NdSbTe.  Moreover, a significant 
hump in the specific heat data in the temperature range of 5-15 K could be 
attributed to the Zeeman splitting \cite{Sankar2020}. The observed broader peak in 
$Cp/T$ versus $T$ data, centered at 48 K, (Fig. \ref{fig4}(b)) indicates 
a Schottky anomaly due to the crystal-field splitting, similar to observed 
in NdSbTe \cite{Sankar2020}. 
The specific heat can be associated with an entropy change in the magnetic phase 
transition. This change in the entropy can be obtained by integrating $\Delta C/T$ 
over a temperature range, where $\Delta C$ is the experimental specific heat except 
the phonon contributions \cite{Salters2023}. As discernible from Fig. \ref{fig4}(c),
the entropy attains a value of $\approx$ 8.34 J/K mol, slightly lower than
the theoretical value of 11.52 J/K mol. The reason 
for this small difference could be due to the approximate analysis and an error 
associated with entropy calculation in milikelvin range. A similar trend in 
entropy is also observed in NdSbTe \cite{Pandey2020} and 
GdSbTe \cite{Sankar2019} materials.

\subsection{Transport and Magnetotransport Properties}

Next, we examined the transport and magnetotransport properties in NdBiTe. For 
this, we performed the temperature-dependent resistivity, $\rho(T)$, 
measurement for polycrystalline pellet of NdBiTe using the standard four-probe method in 
the temperature range of 2 - 300 K. The data from our measurement is shown in 
Fig. \ref{fig5}. As discernible from the panel (a) of the figure, consistent
with the metallic behavior, $\rho$ decreases linearly with a decrease in 
the temperature in the range 20 K $< T <$ 300 K. However, $\rho$ is found 
to increase with further decrease in temperature below 20 K and it could be 
attributed to either Kondo-like behaviour \cite{Barua2017} or electron-electron 
interaction \cite{Li2022}. The observed positive magnetoresistance implies that the 
rise in resistivity originate from electron-electron interaction. Similar
upturn in resistivity at low temperature was observed in GdBiTe \cite{Gebauer2021}.
The presence of magnetic interactions well above the $T_{\rm N}$ observed in 
the heat capacity data could be corroborated with the minima in 
the $\rho(T)$. Similar metallic nature is reported in the case of nonmagnetic
LaSbTe \cite{Singha2017a}, and magnetic GdBiTe \cite{Gebauer2021}, whereas,
other isostructural analogues of early lanthanides Ln = CeSbTe \cite{lv2019}, 
NdSbTe \cite{Pandey2020} and SmSbTe \cite{Pandey2021} are reported to show 
nonmetallic temperature dependence. In addition, nonmonotonic temperature 
dependence of resistivity is reported in HoSbTe\cite{Yang2020} and DySbTe\cite{Gao2022} systems.

The residual resistivity ratio (RRR= R$_{300 K}$/R$_{20 K}$) 
extracted from our measurements is 1.12. Such a small value suggests the 
significant scattering of conducting electrons, which is consistent with 
the other magnetic isostructural analogs with metallic behavior in 
resistivity data \cite{Gebauer2021}. Whereas, the nonmagnetic analogues like LaSbTe \cite{Singha2017a} 
and ZrSiS \cite{Singha2017, Hu2016} are reported to have significantly larger 
RRR values. The $\rho(T)$ shows linear temperature dependence in the range of 
100-300 K and changes its slope at lower temperatures, which can be analyzed more closely
to get information about the phonon scattering. For this, we fit $\rho(T)$ data 
with the Bloch-Gruneisen equation for acoustic electron-phonon scattering, 
in the temperature range (20-100 K), as
\begin{equation}
\begin{split}
  \rho (T) = \rho (T_0) \\
  + A \left(\frac{T}{\Theta_R}\right)^5 
	\int^{\Theta_R/T}_0 \left[ \frac{x^5}{(e^x-1)(1-e^{-x})} \right] dx.
\label{Bloch}
\end{split}
\end{equation}
Here, $\rho (T_0)$ is residual resistivity, the parameter $A$ is a pre-factor 
which depends on the electron-phonon coupling strength, and $\Theta_R$ is the 
Debye temperature derived from resistivity data. From the fit, we obtained the 
value of $A$ to be $1.17(2)\times$ 10$^{-4}$ and $\Theta_R$ 
is extracted to be 150$\pm$2 K, which is close to the $\Theta_D$ obtained 
from specific heat measurement over the same temperature 
range \cite{Jayakumar2021}.

Further, to explore more about the charge carriers and transport mechanisms, 
magnetotransport data were recorded at various temperatures in the presence of 
magnetic field of $\pm$ 90 kOe, applied perpendicular to the current direction.  
The magnetoresistance has been calculated from the relation $MR(\%) = \frac{R(H)- R(0)}{R(0)} \times 100$,
where $R(H)$ and $R(0)$ are resistance at finite applied field and zero field.
As discernible from Fig. \ref{fig5}(b), positive magnetoresistance were 
observed for all the measured temperatures. The magnitudes of MR were found 
to lie in the range of 1 to 2\%. The relatively lower MR values at low 
temperatures infer interacting fermions having less mobility, as observed 
in other AFM lattices \cite{Zhang2017}. Consistent with our data, 
similar low values for MR were observed in LaSbSe \cite{Pandey2022} and 
GdBiTe \cite{Gebauer2021}.
The symmetrical MR shows a change in slope at two applied fields 
$H_1$ and $H_2$. As can be observed from the panel (c) of the figure, 
MR and corresponding $d(MR)/dH$ at 2 K have critical fields as 10 kOe and 
34.5 kOe. As the temperature increases, the lower critical field, which 
is related to the ordered magnetic state, disappears. The $d(MR)/dH$ for 
all temperatures are given in the supplementary material. Further, the correlated 
nature of electrons in NdBiTe can be parameterized with the calculation of 
the Wilson ratio, which is expressed in terms of the low-temperature magnetic 
susceptibility and Sommerfeld coefficient, as 
$R_{\rm w} = \frac{\pi^2 k_{\rm B}^2 \chi} {\mu_{\rm eff}^2 \gamma}$,
where $k_{\rm B}$ is Boltzmann constant, $\chi$ represents magnetic susceptibility,
$\mu_{\rm eff}$ is effective magnetic moment, and $\gamma$ is the Sommerfeld coefficient.
The obtained Wilson ratio is $\approx$ 12.6. Such a large value corroborates 
with correlated fermionic behavior and might have originated from the 
crystal field splitting of the $f$-states.

\subsection{First-Principles Calculations}

Next, to support our experimental data, we performed density functional 
theory based first-principles calculations of the relevant properties in 
NdBiTe. We start with the structural optimization to achieve the actual ground
state of the system. Our optimized lattice parameters, a = b = 4.467 {\AA} 
and c = 9.589 \AA, are in good agreement with the experimental values, 
a = 4.424  {\AA} and c = 9.377 \AA. The optimized unit cell is slightly 
elongated. And, the reason for this could be attributed to the use of 
GGA pseudopotential in our calculations, which is known to 
overestimate the lattice parameters \cite{Stampfl1999}.

\begin{figure}[t!]
\begin{center}
  \includegraphics[scale = 0.35]{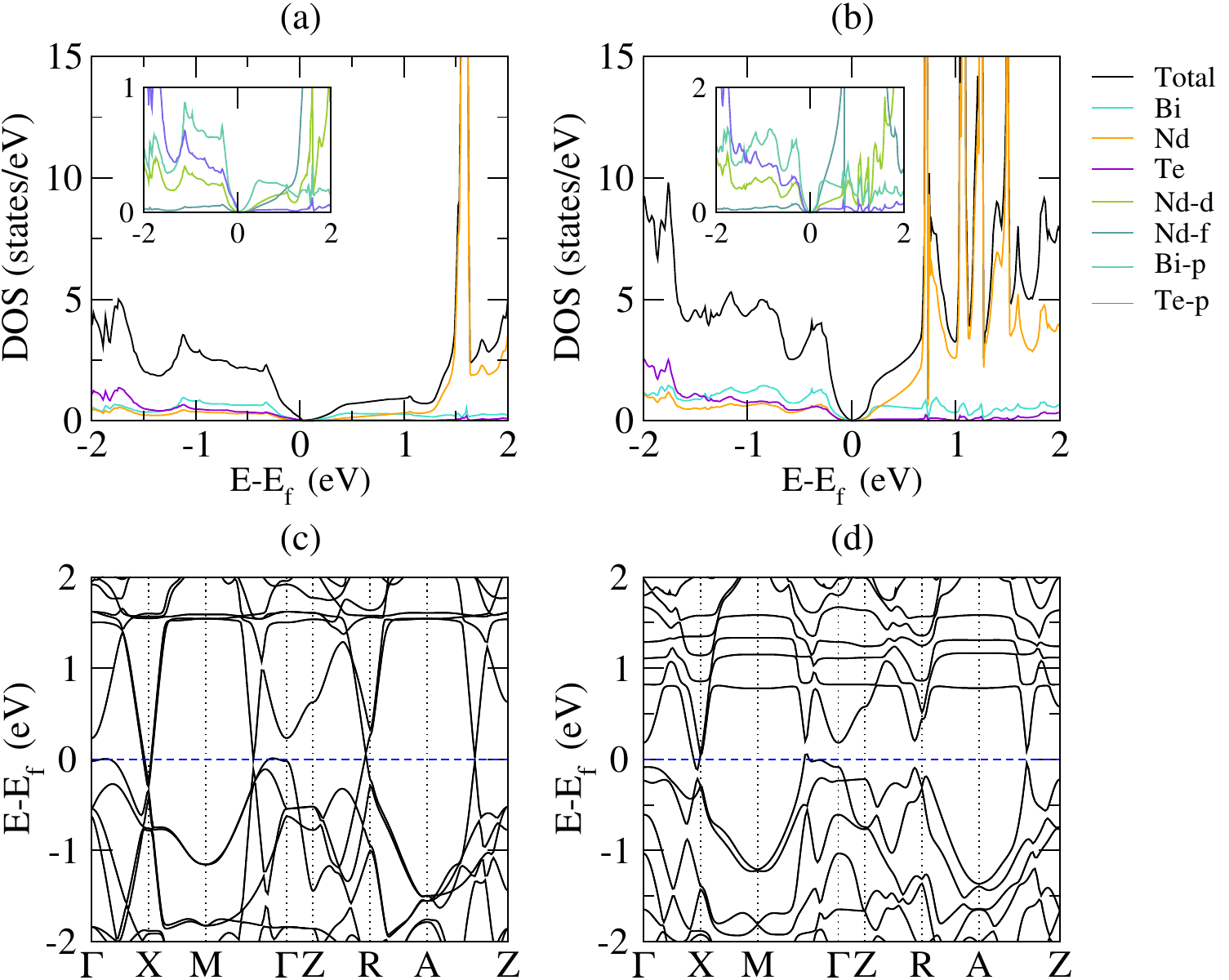}
  \caption{The density of states and atom-projected
        density of states (a) without and (b) with spin-orbit coupling. 
	Insets in panels (a) and (b) represent the orbital resolved
        density of states. The electronic band structure of NdBiTe (c) GGA+U (d) GGA+U+SOC. 
	The Fermi level is set to zero.}
  \label{band}
\end{center}
\end{figure}

The optimized structure is then used to calculate and examine the 
electronic structure of the material. In Fig. \ref{band}, we have shown 
our calculated data for density of states (DOS) and bands for the 
ground state of NdBiTe. As discernible from the figure, our calculated 
data reveals that the NdBiTe is semimetallic in nature. This is consistent 
with the semimetallic 
nature observed in our experiment as well as reported nature in the 
case of GdSbTe \cite{Sankar2019, Hosen2018}, GdBiTe \cite{Gebauer2021}, 
NdSbTe \cite{Sankar2020}, and other materials of the ZrSiS family \cite{Su2018,Pandey2021}.  
In panels (a) and (b) of the figure, we have shown our data on total 
DOS and atom-projected DOS, without and with spin-orbit coupling (SOC), respectively.
As discernible from the DOS, the dominant contributions to valence band 
mainly come from the $p$-electrons of Bi and Te. Apart from these, Nd also 
has a significant contribution, where $d$- and $f$-electrons contribute 
the most. Considering the conduction band, it is mainly dominated by Nd-$f$ 
electrons, specially in the energy range above 0.5 eV.

\begin{table}
	\caption{Calculated total energy (in meV) of FM and AFM-A configurations 
	relative to that of AFM-I configuration, and the magnetic moments 
	(in $\mu_{\rm B}$) of the constituent atoms.}
\centering
\begin{tabular}{lr}
\hline\hline
	$E_{\rm AFM-I}$  & 0 \\
	$E_{\rm FM}$     & -35.98    \\
	$E_{\rm AFM-A}$  & -12.69  \\
\hline
  $\mu^{\rm Nd}$ &  3.059 \\
  $\mu^{\rm Bi}$ &  -0.001 \\
  $\mu^{\rm Te}$ &  -0.019 \\
\hline
$E_{\parallel} - E_{\perp}$  &  0.89 \\
\hline \hline
\end{tabular}
\label{afm_e}
\end{table}

\begin{figure}[t!]
 \includegraphics[width=\linewidth]{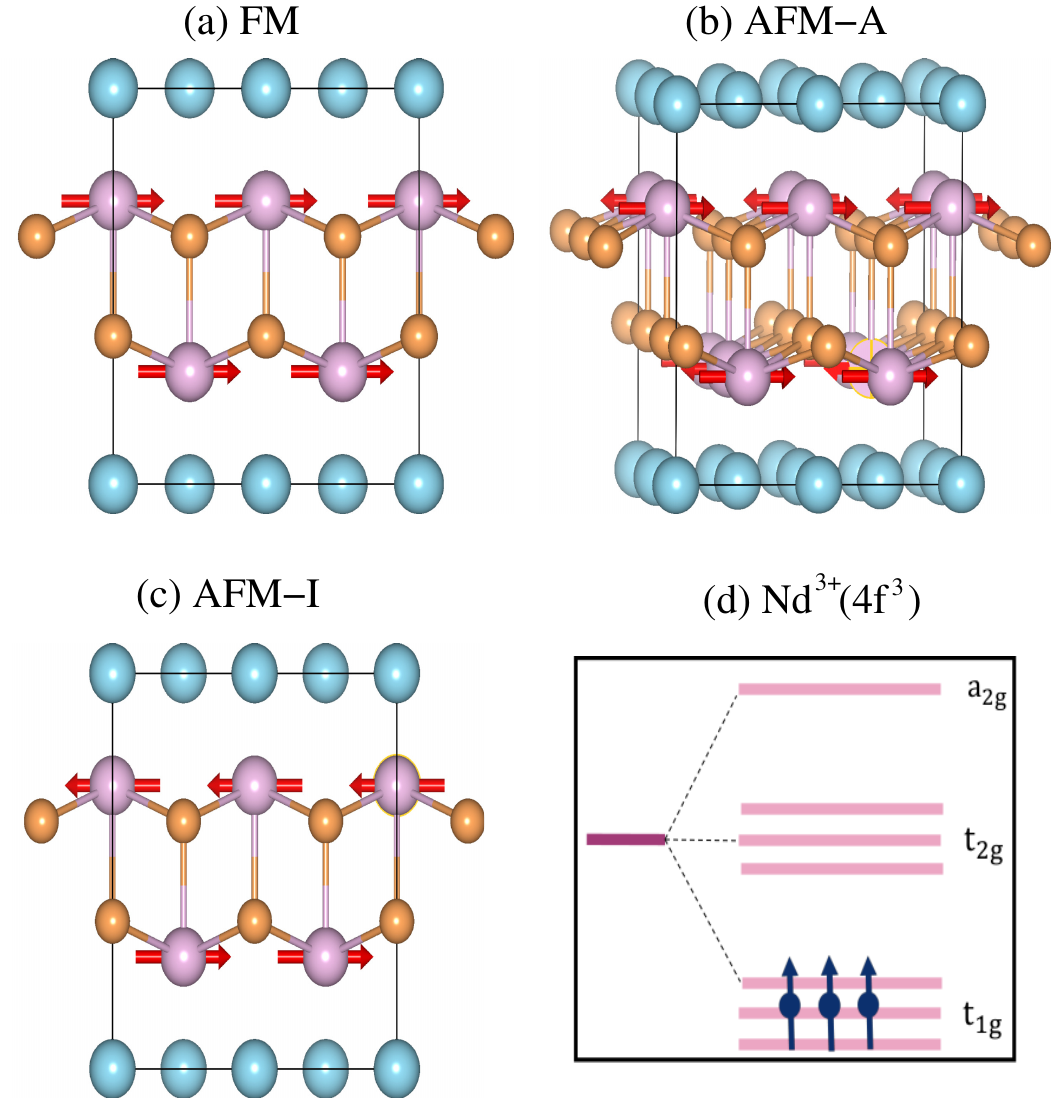}
	\caption{The schematic of (a) FM, (b) AFM-A and (c) AFM-I 
	magnetic configurations. The panel (d) represents the octahedral
	filling of electrons in $f$-state of Nd ion.}
 \label{m_config}
\end{figure}

Next, we examined the magnetic properties of NdBiTe using first-principles 
calculations. First, to obtain the actual ground state magnetic configuration 
of NdBiTe, we probed both ferromagnetic (FM) and antiferromagnetic (AFM) 
orientations of the magnetic moments. For AFM, there 
could be two possible spin configurations -- an interlayer AFM with intralayer 
FM (Fig. \ref{m_config}(c)), and an interlayer FM with intralayer 
AFM (Fig. \ref{m_config}(b)).
In Table \ref{afm_e}, we have listed the total energies for ferromagnetic 
and antiferromagnetic AFM-A with respect to the antiferromagnetic AFM-I configuration. 
As evident from the table, the preferable ground state magnetic configuration of 
NdBiTe is AFM-I, as FM  and AFM-A energies are larger by $\approx$ 35 and 13 meV, respectively.
Our calculation is consistent with our experiment and also the literature \cite{Gebauer2021}  
where a similar isostructural material GdBiTe is reported to have AFM-I \cite{Gebauer2021}
as the ground state. 
 It is to be also mentioned that GdSbTe \cite{Sankar2019} is reported to show an AFM-A magnetic
phase. This implies the importance of interplay between SOC, magnetism, and topological 
degrees of freedom in these materials.
In the table, we have also listed the magnetic moment of 
individual elements. As evident, the dominant contribution is 
from the Nd ion, and the reason for this is attributed to the unpaired $f$-electrons.
As discernible from the octahedral filling of $f$-orbital (Fig. \ref{m_config})(d)), 
the lowest energy state ($t_{1g}$) is occupied by three majority spins, whereas 
the high energy states ($t_{2g}$ and $a_{2g}$) are completely unfilled, leading to 
three unpaired electrons. Our calculated magnetic moment of Nd is 3.06 $\mu_B$ 
per atom, which is consistent with our measured value of 3.64 $\mu_B$. As expected, 
elements Bi and Te have negligible magnetic moments.
Next, to identify the actual direction of spin orientation, we computed the 
magnetocrystalline anisotropy energy (MAE) in NdBiTe. For this, we considered 
three principal spin quantization axes, the out-of-plane [001] and  the in-plane 
[100] and [010]. The MAE energy is extracted from the total energies of in-plane 
and out-of-plane magnetic configurations using $E_{\rm MAE} = E_{\rm in-plane} - E_{\rm out-of-plane}$.
The obtained positive $E_{\rm MAE}$ value, 0.89 meV, indicates that 
NdBiTe prefers an in-plane spin alignment. This is consistent with the previously 
reported $E_{\rm MAE}$ in GdSbTe \cite{Sankar2019}. 

\begin{figure}[t!]
\begin{center}
  \includegraphics[scale = 0.35]{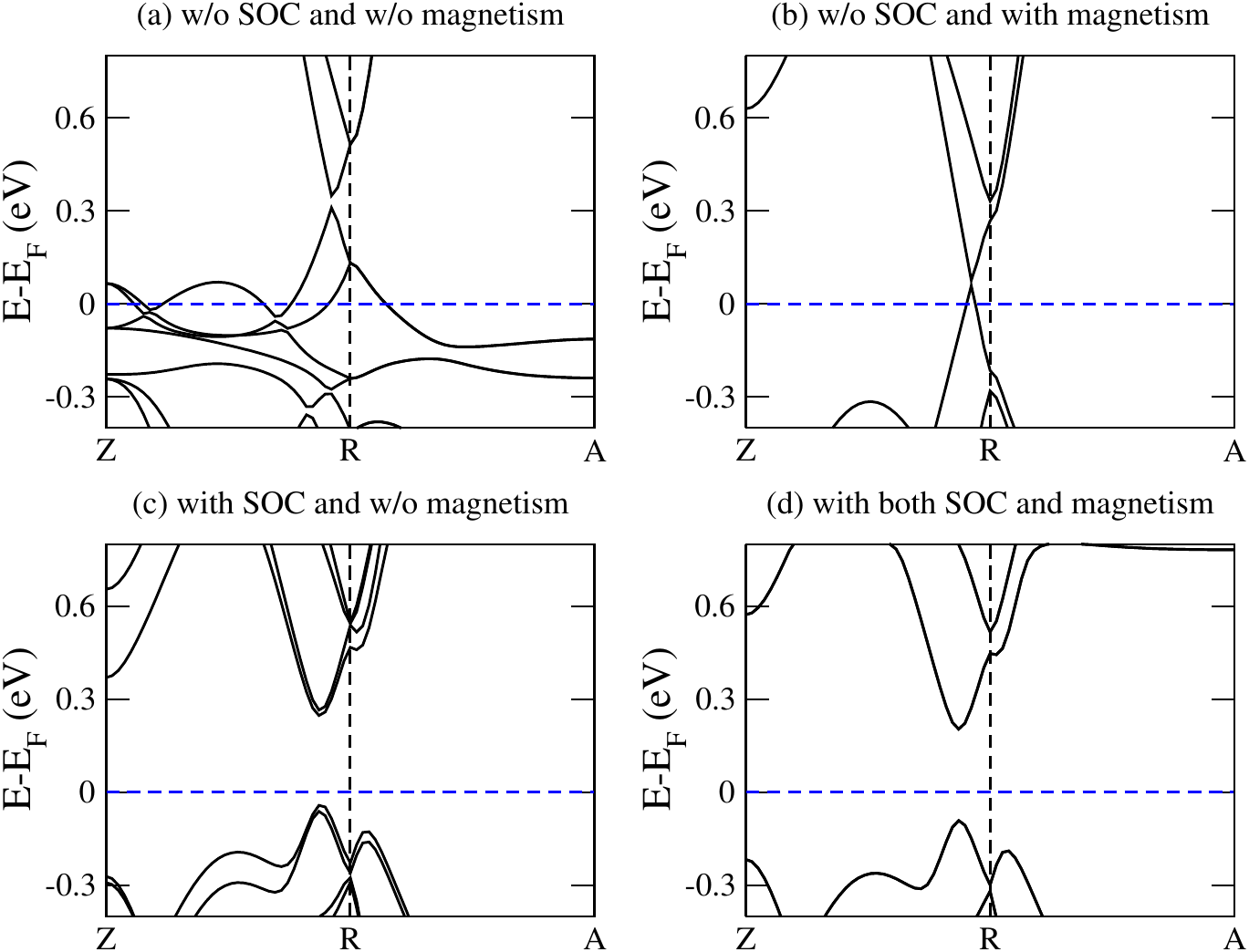}
  \caption{The enlarged view of electronic structure of NdBiTe at high 
	symmetry point R for (a) without SOC and magnetism, (b) without SOC 
	and with magnetism, (c) with SOC and without magnetism, and (b) with 
	SOC and magnetism.}
  \label{R_band}
\end{center}
\end{figure}

\begin{figure}[t!]
\begin{center}
  \includegraphics[width = 8 cm]{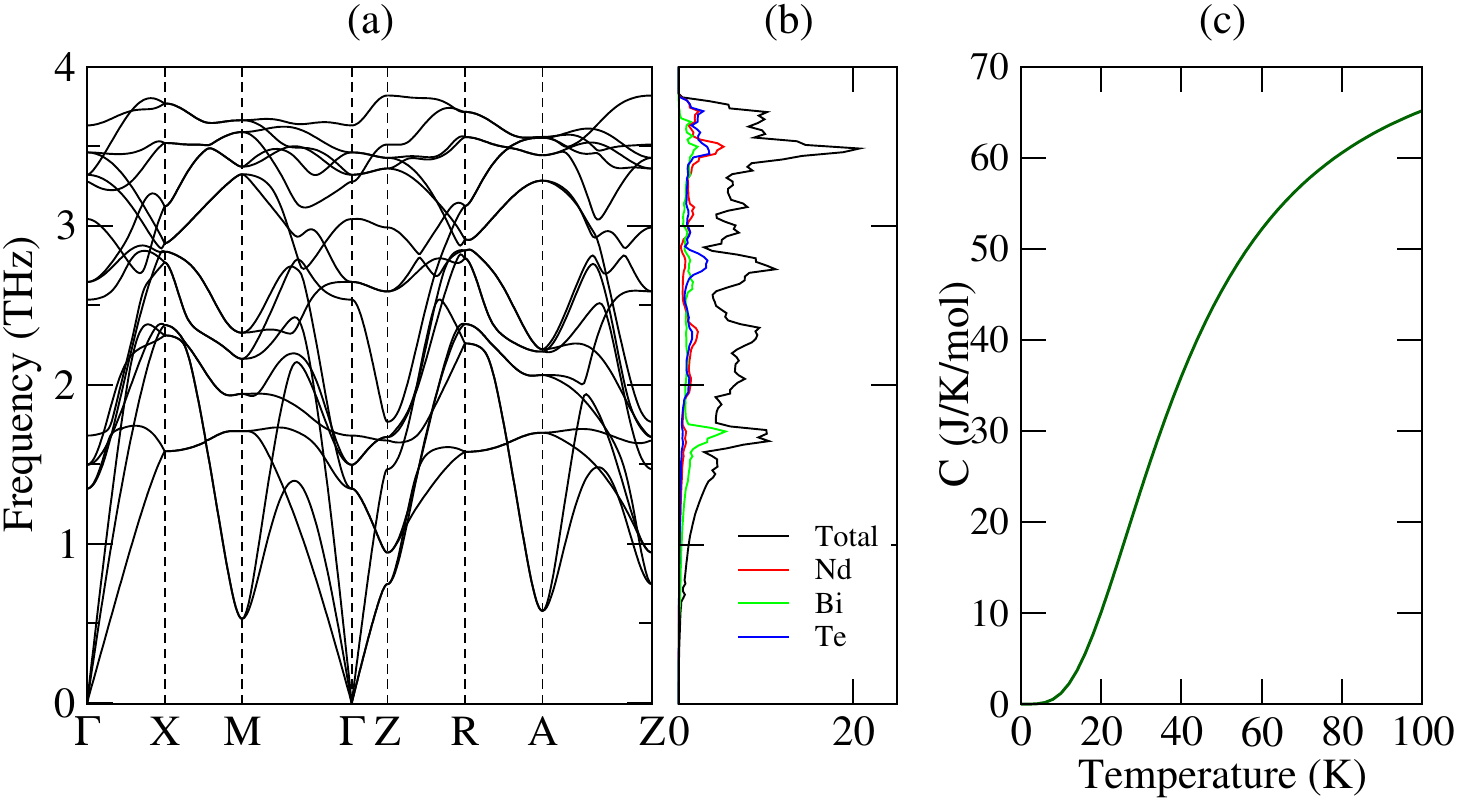}
  \caption{The calculated (a) phonon dispersion (b) phonon density of states 
	and (c) Temperature dependence of specific heat at constant volume 
	for NdBiTe.}
  \label{phonon}
\end{center}
\end{figure}

The ZrSiS-class of materials have a glide-plane perpendicular to a 4-fold axis 
of rotation as a symmetry element due to nonsymmorphic crystalline structure. 
This symmetry facilitates band folding, resulting in band-crossing and, thus, 
nontrivial topology in the material. Considering this, next we examine the 
topological features in NdBiTe. For this, we zoom-in the bands at high-symmetry 
point R and in Fig \ref{R_band} show the enlarge view of bands for four 
different cases. As can be observed from panel (a), in absence of SOC 
and magnetic degrees of freedom, the electronic  structure of NdBiTe resembles
a conventional ZrSiS behavior. However, when the magnetic degrees of freedom 
is switched on, with AFM-I magnetic spin orientation, we observed band crossings, 
leading to nodal lines and nodal surfaces, along the high symmetry directions 
in the Brillouin zone (Fig. \ref{band}(c)). The enlarged view of one such 
crossing is shown in Fig. \ref{R_band}(b) at high-symmetry point R. And, as 
discernible from panel (c), when the SOC is switched-on in absence of magnetic 
degrees of freedom, the nodal lines and nodal surfaces break apart, leading to 
the formation of degenerate Dirac points protected by time-reversal symmetry. 
In presence of both SOC and magnetism, an opening of bands is introduced 
in the material (panel (d)). A similar trend of gap openings in presence of 
SOC were reported recently in HoSbTe \cite{Yang2020}, DySbTe \cite{Gao2022}, 
GdBiTe \cite{Gebauer2021}, CeBiTe, and LaBiTe \cite{Weiland2019,Lei2021} 
ZrSiS class of materials. Such behavior of electronic structure is associated 
with a weak topological nature of the material.

Next, we present and discuss our simulation results on phonon dispersion and 
specific heat in NdBiTe. The phonons were obtained by solving 
the equation (as implemented in the Phonopy \cite{Togo15})
\begin{equation}
  \sum_{\beta\tau'} D^{\alpha\beta}_{\tau \tau'}
        (\mathbf{q}) \gamma^{\beta\tau'}_{\mathbf{q}j} = 
        \omega^2_{\mathbf{q}j}\gamma^{\alpha\tau}_{\mathbf{q}j}, 
\end{equation}
where the indices $\tau, \tau'$ represent the atoms,
$\alpha, \beta$ are the Cartesian coordinates, ${\mathbf{q}}$ is
a wave vector and $j$ is a band index.  $D(\mathbf{q})$ refers to as
the dynamical matrix, and $\omega$ and $\gamma$ are the corresponding
phonon frequency and polarization vector, respectively.
The phonon dispersion and corresponding DOS from our simulations are 
shown in Fig.~\ref{phonon}. As we observed from the panel (a) of the figure, 
the absence of imaginary frequencies confirms the dynamical stability of NdBiTe. 
The acoustic modes were found to be dominated by square-net of Bi, whereas 
the optical phonons have contributions mostly from Nd and Te ions. 
The observed relatively low frequency phonons, compared to ZrSiS/Se \cite{Singha2018, Zhang2017}, 
could be attributed to the presence of heavy elements in NdBiTe.
In panel (c) of the figure, we have shown the temperature dependence of the 
specific heat data, computed within the quasi-harmonic approximation \cite{Baroni2010}.
As can be observed from the figure, our calculated specific heat is consistent 
with our measurement in terms of trend, except an anomaly observed experimentally 
at 4.5 K. The values of $\gamma$ and $\beta$ extracted from the linear 
fit of specific heat data in the experimental temperature range (18 - 26 K) 
are 190 mJ/K$^2$ and 0.76 mJ/K$^4$, respectively. And, $\theta_D$ calculated 
using $\gamma$ and $\beta$ is obtained as 197 K. The reason for the 
small discrepancy with experimental extracted values could be attributed 
to the missing magnetic contributions in our calculation, which are crucial 
in the vicinity of the transition temperature.

\section{CONCLUSIONS}

We have synthesized a polycrystalline NdBiTe by using the sealed tube method 
of solid-state reaction technique. 
From the temperature-dependent magnetic and specific heat measurements, 
we observed an antiferromagnetic configuration as the ground state of NdBiTe, 
with a Neel temperature of 4.5 K. The Neel temperature is observed to be 
less affected with the increase in magnetic field strengths, suggesting the 
robust nature of the phase transition.
In the AFM state, the magnetic field-driven metamagnetic transition is 
observed from the isothermal magnetization data. 
The absence of short-range interactions is confirmed by our 
ac-susceptibility measurements, performed at various excitation frequencies. 
The extracted $\gamma$, $\beta$, and $\theta_{\rm D}$ parameters from 
specific heat data, suggest the presence of relatively stronger electronic 
correlations and weaker interatomic interactions in NdBiTe as compared 
to NdSbTe. Moreover, the transport studies infer the metallic nature with 
positive magnetoresistance. Consistent with our experimental results, our 
first-principles density functional theory based calculations predict an 
antiferromagnetic semimetallic nature of NdBiTe. Our simulations also 
indicate a  weak topological nature associated with NdBiTe. The observed 
topological and magnetic characteristics from our work establish NdBiTe 
as a potential candidate for AFM-based spintronics applications.

\section*{ACKNOWLEDGMENTS}
The authors acknowledge the Central Research Facility (CRF), IIT Delhi 
for the experimental facilities. We thank the MPMS3 facility of the 
Physics department, IIT Delhi for the ac-susceptibility measurements. 
AKG and BKM thanks SERB-DST (RP04583), Government of India for funding.
PKM acknowledges the Council of Scientific $\&$ Industrial Research (CSIR), 
[09/086(1425)/2019-EMR-I] India for fellowship. BKM acknowledges the 
funding support from SERB, DST (CRG/2022/000178). The results presented 
in the paper are based on the computations using the High Performance 
Computing cluster, Padum, at the Indian Institute of Technology Delhi, New Delhi

\begin{suppinfo}

The supplementary information includes isothermal magnetization $M(H)$ plot and the first derivative of MR with respect to the applied field at various temperatures. It also consists of U-value calculation for Nd, utilized further in theoretical investigations.

\end{suppinfo}


\providecommand{\latin}[1]{#1}
\makeatletter
\providecommand{\doi}
  {\begingroup\let\do\@makeother\dospecials
  \catcode`\{=1 \catcode`\}=2 \doi@aux}
\providecommand{\doi@aux}[1]{\endgroup\texttt{#1}}
\makeatother
\providecommand*\mcitethebibliography{\thebibliography}
\csname @ifundefined\endcsname{endmcitethebibliography}  {\let\endmcitethebibliography\endthebibliography}{}

\end{document}